\documentclass[preprint,12pt]{elsarticle}

\usepackage{amssymb}
\journal{Physics Letters B}

\begin{document}

\begin{frontmatter}

%% Title, authors and addresses

%% use the tnoteref command within \title for footnotes;
%% use the tnotetext command for theassociated footnote;
%% use the fnref command within \author or \address for footnotes;
%% use the fntext command for theassociated footnote;
%% use the corref command within \author for corresponding author footnotes;
%% use the cortext command for theassociated footnote;
%% use the ead command for the email address,
%% and the form \ead[url] for the home page:
%% \title{Title\tnoteref{label1}}
%% \tnotetext[label1]{}
%% \author{Name\corref{cor1}\fnref{label2}}
%% \ead{email address}
%% \ead[url]{home page}
%% \fntext[label2]{}
%% \cortext[cor1]{}
%% \address{Address\fnref{label3}}
%% \fntext[label3]{}

\title{Half-lives of $\alpha$ decay from natural nuclides and from superheavy elements}

%% use optional labels to link authors explicitly to addresses:
%% \author[label1,label2]{}
%% \address[label1]{}
%% \address[label2]{}

\author[a,b]{Yibin~Qian \corref{cor}} \ead{qyibin@gmail.com}
\author[a,b,c,d]{Zhongzhou~Ren \corref{cor}}
\cortext[cor]{Corresponding author.} \ead{zren@nju.edu.cn}
\address[a]{Key Laboratory of Modern Acoustics and Department of
Physics, Nanjing University, Nanjing 210093, China}
\address[b]{Joint Center of Nuclear Science and Technology, Nanjing
University, Nanjing 210093, China}
\address[c]{Kavli Institute for Theoretical Physics China, Beijing 100190, China}
\address[d]{Center of Theoretical Nuclear Physics, National Laboratory of Heavy-Ion
Accelerator, Lanzhou 730000, China}

\begin{abstract}
Recently, experimental researches on the $\alpha$ decay with long
lifetime are one of hot topics in the contemporary nuclear physics
[e.g. N. Kinoshita {\sl et al.} (2012) \cite{Sm146} and J. W. Beeman
{\sl et al.} (2012) \cite{Bee1}]. In this study, we have
systematically investigated the extremely long-lived
$\alpha$-decaying nuclei within a generalized density-dependent
cluster model involving the experimental nuclear charge radii. In
detail, the important density distribution of daughter nuclei is
deduced from the corresponding experimental charge radii, leading to
an improved $\alpha$-core potential in the quantum tunneling
calculation of $\alpha$-decay width. Besides the excellent agreement
between theory and experiment, predictions on half-lives of possible
candidates for natural $\alpha$ emitters are made for future
experimental detections. In addition, the recently confirmed
$\alpha$-decay chain from $^{294}$117 is well described, including
the attractive long-lived $\alpha$-decaying $^{270}$Db, i.e., a
positive step towards the ``island of stability'' in the superheavy
mass region.

\end{abstract}

\begin{keyword}
\PACS 23.60.+e \sep 21.60.Gx \sep 21.10.Ft \sep 21.10.Tg
%% keywords here, in the form: keyword \sep keyword

%% PACS codes here, in the form: \PACS code \sep code

%% MSC codes here, in the form: \MSC code \sep code
%% or \MSC[2008] code \sep code (2000 is the default)

\end{keyword}

\end{frontmatter}

%% \linenumbers

%% main text
\section{Introduction}

Since the discovery of radioactive decays in the 1890s \cite{Bec},
$\alpha$ decay has always played a quite important role in both the
foundation and development of nuclear physics. In the recent
experimental studies, one hot subject was to detect the naturally
long-lived $\alpha$-decaying nuclides. A shorter $\alpha$ decay
half-life of $^{146}$Sm was recently measured, which appears to be
quite valuable due to the significance of $^{146}$Sm--$^{142}$Nd
(its $\alpha$-decay daughter) chronology in the solar system
\cite{Sm146}. For a long time, the naturally occurring $^{209}$Bi
was believed to be the heaviest stable nuclide until the observation
of its $\alpha$-decay by Marcillac {\sl et al.} \cite{Marc} and the
first measurement of the partial widths by J. W. Beeman {\sl et al.}
\cite{Bee1}. Lead has then been supposed to be the heaviest stable
element, and new experimental limits were newly proposed for the
$\alpha$ decays of Pb isotopes \cite{Bee2}. In fact, the detection
on natural $\alpha$ radioactivity can be dated back to 1960s
\cite{Macf}, and it has received much more attention with the
development of the facilities. Besides the above mentioned cases,
much effort has been made for probing the rare $\alpha$ activity of
$^{180}$W \cite{Dane,Cozz}, a series of experiments were performed
on the $\alpha$ decay of natural europium \cite{Belli,Casali}, etc.
Additionally, it is exciting that the researchers have independently
confirmed the existence of new element 117 \cite{Khuya} since the
original experiment at Dubna in 2010 \cite{Yu}, which actually marks
the official status of this new element. This newly discovered
$\alpha$ decay chain from $^{294}$117 to a new isotope $^{266}$Lr
even includes a hitherto longest-lived $\alpha$-emitters $^{270}$Db
among heaviest elements, indicating a possible milestone towards the
location of the ``island of stability''. It is of great physical
interest to pay attention to various long-lived $\alpha$ emitters in
nature, and the striking $\alpha$ decay chain populating the
superheavy nucleus with long lifetime.

Following the quantum explanation of $\alpha$ decay by Gamow in 1928
\cite{Gamow}, $\alpha$ decay is usually considered as the tunneling
process of the preformed $\alpha$ particle through the barrier
potential. With the help of phenomenological and effective
$\alpha$-core potentials, theoretical studies
\cite{Buck1,Lovas,Denisov1,Denisov2,Royer1,Royer2,Silis,Santhosh,Xu,Mohr,DN,Ren,Ni1,Reny,Ismail,Seif,Qian1,Qian2,Qian3}
have been subsequently proposed for $\alpha$ decay calculations
especially in the last two decades. Among these studies, our group
provided a unified formula for half-lives of $\alpha$ decay and
cluster radioactivity \cite{Ni1} and a new Geiger-Nuttall relation
was recently proposed for $\alpha$ decay including the effects of
the quantum numbers of $\alpha$-core relative motion \cite{Reny}. In
this Letter, we present a generalized density-dependent cluster
model to depict the attractive naturally occurring $\alpha$
emissions, involving the density distributions of residual daughter
nuclei based on their experimental root-mean-square (rms) charge
radii. Fortunately, there are generally available experimental
charge radii for these focused daughter nuclei \cite{Angeli}. After
the total $\alpha$-core potential is constructed via the
double-folding procedure combined with the effective M3Y-Reid-type
nucleon-nucleon (NN) interaction and the standard Coulomb
proton-proton interaction, the tunneling calculation is simplified
as a bound state problem and a scattering state problem according to
the modified two-potential theory \cite{Gurvitz}. The eigen
characteristic of the bound state for the $\alpha$ particle is
determined approximately by the Wildermuth condition
\cite{Wildermuth}, which relates the quantum numbers of the $\alpha$
cluster to the shell-model quantum numbers of the nucleons forming
the cluster. This in fact takes into account the main requirement of
the Pauli exclusion principle, and the remaining effects are
absorbed into the fitting parameters of the effective
$\alpha$-nucleus potentials.

\section{Theoretical framework}

Given the assumption that an $\alpha$ cluster interacts with an
axially symmetric deformed core nucleus, the total interaction
potential of the $\alpha$-core system comprises of the nuclear and
Coulomb potentials plus the centrifugal term,
\begin{eqnarray}
V(r,\theta)~=~\lambda
V_N(r,\theta)+V_C(r,\theta)+\frac{\hbar^2\ell(\ell+1)}{2\mu r^2},
\end{eqnarray}
where $\lambda$ is the renormalization factor for nuclear potential,
$\theta$ is the orientational angle of the emitted $\alpha$ particle
with respect to the symmetric axis of the daughter nucleus, $\mu$ is
the reduced mass of the $\alpha$-daughter system in the unit of the
nucleon mass $\mu=A_{\alpha}A_d/(A_{\alpha}+A_d)$, and $\ell$ is the
angular momentum carried by the $\alpha$ cluster. In the
density-dependent cluster model, the nuclear and Coulomb potentials
are obtained by the double-folding integral of the realistic NN
interaction with the density distributions of the $\alpha$ particle
and the residual core nucleus \cite{Bertsch,Satchler},
\begin{eqnarray}
V_{N or C}(\mathbf{r},\theta)~=~\int\int
d\mathbf{r_1}~d\mathbf{r_2}~\rho_1(\mathbf{r_1})\upsilon(\mathbf{s=|r_2+r-r_1|})\rho_2(\mathbf{r_2}),\label{potential}
\end{eqnarray}
where $\upsilon(\mathbf{s})$ denotes the widely-used M3Y NN
interaction derived from the $G$-matrix elements of the Reid
potential for the nuclear potential \cite{Bertsch}. When the above
formula serves for the Coulomb component, the $\upsilon(\mathbf{s})$
represents the standard Coulomb proton-proton interaction. The
density distribution of the spherical $\alpha$ particle is the
standard Gaussian form given in the high-energy electron scattering
experiment \cite{Satchler}. On the other hand, in contrast with the
available information on experimental nuclear charge radii, the
nuclear neutron distribution appears to be extremely ambiguous.
Subsequently, the specific formula for the charge distribution can
be approximately obtained based on the experimental detection, as
described in the following. It is hard to analytically depict the
neutron distribution from the poor knowledge of nuclear neutron
radii or neutron skin thickness in nuclei \cite{Ni2}. Considering
this, we assume that the density distribution of neutrons has the
same form with that of protons in nuclei, i.e., the mass and charge
density distributions of the daughter nucleus are both supposed to
behave in the Fermi form,
\begin{eqnarray}
\rho_2(r_2,\theta_1)~=\frac{\rho_0}{1+\exp\Big[\frac{r_2-R(\theta_1)}{a}\Big]}.\label{rho}
\end{eqnarray}
Here the half-density radius $R(\theta_1)$ is parameterized as
$R(\theta_1)=r_0A_d^{1/3}[1+\beta_2Y_{20}(\theta_1)+\beta_4Y_{40}(\theta_1)]$,
and $a$ is the diffuseness parameter. The $\rho_0$ value is
determined by integrating the density distribution equivalent to the
mass or atomic number of the residual daughter nucleus, and the
quadrupole ($\beta_2$) and hexadecapole ($\beta_4$) deformation
parameters are taken from the theoretical values given by M\"{o}ller
{\sl et al.} \cite{Moller}. The $\alpha$-core potential can then be
obtained by the double-folding integral of the effective NN
interaction with the aforementioned density distributions, within
the multipole expansion method (see details in Refs.
\cite{Xu,Satchler} and references therein). Given one certain angle
$\theta$, the total potential $V(r,\theta)$ is reduced into one
dimensional case, namely $V(r)$. Within the two-potential approach,
$V(r)$ is then divided into two parts: the ``inner'' term and the
``outer'' term by a separation radius, and the Schr\"{o}dinger
equation is numerically solved in the inner potential for the bound
state wave function. Because the decay energy $Q$ is very sensitive
to the half-life calculation and it cannot be predicted with
sufficient accuracy for a given potential as well, we adjust the
$\lambda$ factor to the experimental $Q$ value for each decay.
Meanwhile, to reflect the Pauli exclusion principle, the quantum
number $n$ of the bound solution (i.e., the number of internal
nodes) is chosen by the Wildermuth condition \cite{Wildermuth},
\begin{eqnarray}
G~=~2n+\ell~=~\sum_{i=1}^{4}g_{i}.
\end{eqnarray}
In this expression, $g_i$ are the corresponding oscillator quantum
numbers of the ingredient nucleons in the $\alpha$ cluster, whose
values are restricted to ensure the $\alpha$ cluster completely
outside the shell occupied by the core nucleus. Here we take $g_i=$
4 for nucleons with $50\leq Z,N\leq82$, $g_i=$ 5 for nucleons with
$82< Z,N\leq126$, and $g_i=$ 6 for nucleons beyond the $N=$ 126
neutron shell closure. Moreover, a zero-range term for the
single-nucleon exchange is introduced in the M3Y NN interaction to
guarantee the antisymmetrization of identical nucleons in the
$\alpha$ cluster and in the core nucleus \cite{Satchler}.
Subsequently, one can use the wave function to obtain the $\alpha$
decay width $\Gamma(\theta)$ for the given angle, as described in
previous studies \cite{Qian1,Qian2}. By averaging the width in
various directions \cite{Denisov1,Denisov2,Xu}, the final decay
width is given by
\begin{eqnarray}
\Gamma=\int_{0}^{\pi/2}\Gamma(\theta)\sin(\theta)d\theta.
\end{eqnarray}

Previously, the parameters $r_0$ and $a$ in the density distribution
are suggested at $r_0=$ 1.07 fm and $a=$ 0.54 fm from the nuclear
textbook \cite{Walecka}, which could lead to the calculated decay
width. In the present Letter, we make use of the corresponding
experimental nuclear charge radii to determine the related
parameters to pursue a better description of the naturally occurring
$\alpha$ activities with long half-lives. In detail, the $\alpha$
particle in the decay process is usually considered to be formed in
nuclear surface, which seems to be directly related with the
half-density radius, namely $r_0$ factor. Importantly, besides the
intuitive knowledge, we found that the final decay width is more
sensitive to the quantity $r_0$ as compared to the diffuseness
parameter $a$ \cite{Ni2}. On the other hand, the focused natural
$\alpha$ emitters are generally in the medium mass region of nuclide
chart. While the diffuseness value ($a=$ 0.54 fm) is suitable for
the $\alpha$ decay studies in heavy nuclei \cite{Xu,Ni2}, the $a$
values in the density distribution should be relatively less ones
for medium nuclei. Based on these facts, the diffuseness $a$ value
is fixed at the following constants: $a=$ 0.54 fm for heavy nuclei
with $N>$126, and $a=$ 0.52 fm for medium nuclei with $N\leq$ 126.
$r_0$ is then considered as the representation factor of the rms
nuclear charge radii. The $r_0$ value of $\rho_2$ for daughter
nuclei can be conveniently obtained from the experimental charge
radii by the relationship
\begin{eqnarray}
R\equiv\sqrt{<r^2>}=\Big[\frac{\int\rho_2(r,\theta_1)r^4\sin\theta_1
dr d\theta_1}{\int\rho_2(r,\theta_1)r^2\sin\theta_1 dr
d\theta_1}\Big]^{1/2}.\label{radius}
\end{eqnarray}

After the decay width is proceeded through the above sequential
procedure, the $\alpha$ decay half-life is ultimately related as
\begin{eqnarray}
T_{1/2}~=~\frac{\hbar \ln2}{P_{\alpha}\Gamma},
\end{eqnarray}
where the $\alpha$-preformation factor $P_{\alpha}$ inscribes the
preformation probability of an $\alpha$ cluster in the parent
nucleus. Its value can, in principle, be evaluated from the overlap
between the actual wave function of the parent nucleus and that of
the decaying state depicting the $\alpha$ cluster coupled to the
residual daughter nucleus. However, it is in fact extremely
difficult to achieve these wave functions due to the complexity of
both the nuclear potential and the nuclear many-body problem.
According to the experimental analysis, the preformation factor
should vary smoothly in the open-shell region and has a value less
than unity \cite{Hodgson}. Considering this, the
$\alpha$-preformation factor is taken as the same constant for one
kind of nuclei, to keep the minimum of free parameters in the model
as well. Through a least square fit to the experimental half-lives
of those long-lived $\alpha$ emitters, the $P_{\alpha}$ values are
chosen as: $P_{\alpha}^{e-e}=$ 0.42 for even-even nuclei and
$P_{\alpha}^{odd-A}=$ 0.15 for odd-$A$ nuclei. This is consistent
with the Buck's model \cite{Buck1}, and these values are close to
the microscopic calculation of the typical nucleus $^{212}$Po
\cite{Lovas}. There is no doubt that the experimental $\alpha$ decay
half-lives should be better reproduced if the preformation factor is
considered as a variable along with different parent nuclei instead
of a constant. Several detections have been performed for this
subject, especially for the closed-shell nuclei
\cite{Ismail,Seif,Qian1,Qian2,Qian4}. This deserves further
investigation.

\section{Numerical results and discussions}

We initially pay main attention to the long-lived $\alpha$-decaying
nuclides in nature, within a generalized density-dependent cluster
model as described above. Table~\ref{tab1} presents our calculated
results for the $\alpha$ decay properties of these focused emitters,
which generally decay from ground states to ground states as listed
in the first column. The next two columns list the experimental
decay energies $Q$ and half-lives, which are taken from the AME2012
\cite{AME}, the NNDC \cite{NNDC} databases, and the newly detected
data within improved accuracy \cite{Sm146,Marc,Bee1,Bee2,Casali}.
The fourth and fifth columns denote the experimental charge radii of
daughter nuclei \cite{Angeli} and the extracted $r_0$ values in the
density distribution [Eq.~(\ref{rho})], respectively. Additionally,
the renormalization factor $\lambda$, namely another important
quantity, is determined in the aforementioned calculation process
and actually varies in a small range of 0.613-0.707. The present
calculated results are given in the sixth column. In detail, these
$\alpha$ decays usually choose the favored ones with $\ell=$ 0 on
the basis of the available experimental assignments, except for
$^{151}$Eu and $^{209}$Bi. According to the new discovery
\cite{Belli,Casali}, the spin and parity of $^{151}$Eu and
$^{147}$Pm are respectively assigned as 5/2$^{+}$ and 7/2$^{+}$,
leading to the minimum $\ell$ = 2 following the spin-parity
selection rule. It should also be noted that the nuclear charge
radius of $^{147}$Pm is taken from the estimation and systematics of
the isotopic chain due to its absence in experiment \cite{Angeli}.
For the $\alpha$ decay of $^{209}$Bi, the angular momentum
transferred by the cluster should be $\ell$ = 5, resulting from the
decay scheme 9/2$^{-}\rightarrow$1/2$^{+}$ \cite{Marc,Bee1}.
Simultaneously, the transition from $^{209}$Bi is strongly effected
by the neutron closed-shell ($N=$ 126) and its $P_{\alpha}$ value
has to be exclusively chosen as the same one proposed in our
previous studies on exotic $\alpha$ decays \cite{Qian1}. Moreover,
the last two columns in Table~\ref{tab1} respectively list the
results obtained by the united model for $\alpha$ decay and $\alpha$
capture UMADAC (Ref.~\cite{Denisov2} and codes therein), and the
analytic expressions for $\alpha$ decay half-lives for full set of
nuclei \cite{Royer2}, to preform the comparison of the present
approach with other ones.

Generally, it is found that our calculated half-lives well agree
with the experimental data within a mean factor of 1.5, and are
comparable to the values given by some other models. Especially, our
calculations are very close to experiments performed for the
important clock $^{146}$Sm in the solar system, the very newly
detected $^{151}$Eu, the unexpected $\alpha$-emitter $^{209}$Bi,
etc. This may imply the significance of considering the sensitive
quantity $r_{0}$ from the nuclear charge radii of daughter nuclei
for computing $\alpha$ decay half-lives. Encouraged by this, we have
also provided predictions on $\alpha$ decay half-lives for
candidates of naturally long-lived $\alpha$ emitters. The rare
$\alpha$ radioactivities of $^{149}$Sm, $^{174,176,178}$Hf,
$^{184}$Os and $^{192}$Pt are strongly suggested for future
experimental researches, in view of their appropriate predicted
half-lives. For example, there is still the uncertainty for the
decay energy of $^{174}$Hf \cite{AME,NNDC}, and its experiment can
serve for the isotopes as well.

As an additional test, we have performed an investigation (listed in
Table~\ref{tab2}) on the decay properties of the $\alpha$ decay
chain originated from $^{294}$117, newly observed in the important
experiment confirming the existence of the new element. The $\alpha$
decay chains from this newly discovered element 117 have in fact
received a lot of attention in theoretical studies (see
Refs.~\cite{Royer2,Silis,Santhosh,Qian3} and references therein)
based on different models such as the shell model \cite{Silis} and
the Coulomb and proximity potential model for deformed nuclei
\cite{Santhosh}, in which the calculated results are generally
consistent with each other. One can see that the new experimental
data \cite{Khuya} exactly provide an opportunity to further check
the validity of the theoretical model. Unfortunately, there is
little knowledge of the level scheme and unavailable information on
the nuclear charge radii of nuclei in the superheavy mass region.
Hence we assume that the $\alpha$ transitions are favored (namely,
$\ell=$ 0) for these superheavy nuclei, and the key parameter $r_0$
and the diffuseness $a$ in their density-distributions are still
fixed at the standard values proposed in the textbook
\cite{Xu,Walecka}, as mentioned before. The $\alpha$-preformation
factor is then taken as the same choice with our previous studies
\cite{Qian2,Qian3} for this case. Furthermore, there are two
possible values of decay energies for $^{282}$Rg in the new
measurement, and we have chosen the identical one with the previous
experiment \cite{Yu}.

With these above in mind, we have calculated the $\alpha$ decay
half-lives by using the experimental decay energies \cite{Khuya}.
Due to the rare events of experiments, there are slightly large
error bars of decay energies for the $\alpha$ decay chain from
$^{294}$117. On one hand, these experimental $Q$ values are
compatible with those in the AME2012 tables \cite{AME}. Some
deviations of them may result from the reason that the mass tables
offer the decay energies between ground states while the
measurements are proceeded for the transitions from or to low
exciting states. On the other hand, the calculated half-live depends
strongly on the decay energy $Q$. There are usually discrepancies of
$Q$ values in various experimental works and theoretical mass tables
especially for superheavy nuclei, which correspondingly bring
different half-lives in extensive theoretical studies. Despite this,
the present calculation and conclusion are not affected. As one can
see from Table~\ref{tab2}, our calculated half-lives, located in a
certain range, slightly underestimate the corresponding experimental
values. This appears to be reasonable because of the possible
disregard for unfavored cases ($\ell\neq$0), which could increase
the calculated half-lives. For the abnormal discrepancy in
$^{286}$113, we may need the enough experimental recognition of
nuclear deformation, energy level and nuclear radius to improve the
calculated result. However, the consistency of the calculation with
the measurement is well reached for these isotopes including the
long-lived $\alpha$-decaying $^{270}$Db, which may be an important
nucleus towards the ``island of stability'' in the heaviest regime.
The following work about the predictions on the attractive ``island
of stability'' is being in the process.

\section{Summary}

In summary, we have developed the density-dependent cluster model to
carefully investigate the naturally $\alpha$-decaying nuclei with
long lifetimes, especially for the newly discovered nuclides and the
important $\alpha$-emitters with improved measurements. The
sensitive parameter in the density distribution of daughter nuclei
in the decay process is obtained from the experimental nuclear
charge radius, in order to pursue an enhanced description of
$\alpha$-core potential via the double-folding method. Our
sequential calculations give the theoretical results of $\alpha$
decay half-lives, excellently agreeing with the experimental data
and being comparable with other theoretical values. As well, we have
made a series of predictions on long half-lives for possible
$\alpha$-decaying nuclei in nature, to be strongly suggested for
future detections. The reconfirming $\alpha$ decay chain from the
new element 117 have also been focused on, to actually further check
the validity of our model to some extent.

\begin{center}
{\large Acknowledgments }
\end{center}
We thank Dongdong Ni and Renli Xu for carefully reading the
manuscript. This work is supported by the National Natural Science
Foundation of China (Grants No. 11035001, No. 10975072, No.
10735010, No. 11375086, and No. 11120101005), by the 973 National
Major State Basic Research and Development of China (Grants No.
2010CB327803 and No. 2013CB834400), by CAS Knowledge Innovation
Project No. KJCX2-SW-N02, by Research Fund of Doctoral Point (RFDP),
Grant No. 20100091110028, and by the Project Funded by the Priority
Academic Programme Development of JiangSu Higher Education
Institutions (PAPD). \vspace{0.5cm}

%% The Appendices part is started with the command \appendix;
%% appendix sections are then done as normal sections
%% \appendix

%% \section{}
%% \label{}

%% If you have bibdatabase file and want bibtex to generate the
%% bibitems, please use
%%
%%  \bibliographystyle{elsarticle-num}
%%  \bibliography{<your bibdatabase>}

\begin{thebibliography}{00}

%% \bibitem{label}
%% Text of bibliographic item

\bibitem{Bec} H. Becquerel, CR Acad. Sci 122 (1896) 501.

\bibitem{Sm146} N. Kinoshita {\sl et al.}, Science 335 (2012) 1614.

\bibitem{Marc} P. D. Marcillac, N. Coron, G. Dambier, J. Leblanc, and J.
Moalic, Nature (London) 422 (2003) 876.

\bibitem{Bee1} J. W. Beeman {\sl et al.}, Phys. Rev. Lett. 108 (2012) 062501.

\bibitem{Bee2} J. W. Beeman {\sl et al.}, Eur. Phys. J. A 49 (2013) 50.

\bibitem{Macf} R. D. Macfarlane, T. P. Kohman, Phys. Rev. 121 (1961) 1758.

\bibitem{Dane} F. A. Danevich {\sl et al.}, Phys. Rev. C 67 (2003) 014310.

\bibitem{Cozz} C. Cozzini {\sl et al.}, Phys. Rev. C 70 (2004) 064606.

\bibitem{Belli} P. Belli {\sl et al.}, Nucl. Phys. A 789 (2007) 15.

\bibitem{Casali} N. Casali {\sl et al.}, J. Phys. G: Nucl. Part. Phys. 41 (2014) 075101.

\bibitem{Khuya} J. Khuyagbaatar {\sl et al.}, Phys. Rev. Lett. 112 (2014) 172501.

\bibitem{Yu} Yu. Ts. Oganessian {\sl et al.}, Phys. Rev. Lett. 104 (2010) 142502;
             \\Yu. Ts. Oganessian {\sl et al.}, Phys. Rev. C 83 (2011) 054315.

\bibitem{Gamow} G. Gamow, Z. Phys. 51 (1928) 204.

\bibitem{Buck1} B. Buck, A. C. Merchant, S. M. Perez, Phys. Rev. Lett. 65 (1990) 2975;
                \\B. Buck, A. C. Merchant, S. M. Perez, At. Data Nucl. Data Tables 54 (1993) 53.

\bibitem{Lovas} R. G. Lovas, R. J. Liotta, A. Insolia, K. Varga, D. S. Delion, Phys. Rep. 294 (1998) 265.

\bibitem{Denisov1} V. Yu. Denisov, H. Ikezoe, Phys. Rev. C 72 (2005) 064613.

\bibitem{Denisov2} KINR UMADAC Code: http://www.nndc.bnl.gov/codes/UMADAC/; V. Yu. Denisov, A. A. Khudenko,
At. Data Nucl. Data Tables 95 (2009) 815; At. Data Nucl. Data Tables
97 (2011) 187 (E).

\bibitem{Royer1} G. Royer, J. Phys. G: Nucl. Part. Phys. 26 (2000) 1149.

\bibitem{Royer2} G. Royer, Nucl. Phys. A 848 (2010) 279.

\bibitem{Silis} I. Silisteanu, A. I. Budaca, At. Data Nucl. Data Tables 98 (2012) 1096.

\bibitem{Santhosh} K. P. Santhosh, B. Priyanka, M. S. Unnikrishnan, Phys. Rev. C  85 (2012) 034604.

\bibitem{Xu} Chang Xu, Zhongzhou Ren, Nucl. Phys. A 760 (2005) 303;
             \\Chang Xu, Zhongzhou Ren, Phys. Rev. C 73 (2006) 041301(R).

\bibitem{Mohr} P. Mohr, Phys. Rev. C 73 (2006) 031301(R).

\bibitem{DN} D. N. Poenaru, I. H. Plonski, W. Greiner, Phys. Rev. C 74 (2006) 014312.

\bibitem{Ren} Zhongzhou Ren, Chang Xu, Zaijun Wang, Phys. Rev. C 70 (2004) 034304.

\bibitem{Ni1} Dongdong Ni, Zhongzhou Ren, Tiekuang Dong, Chang Xu, Phys. Rev. C 78 (2008) 044310.

\bibitem{Reny} Yuejiao Ren, Zhongzhou Ren, Phys. Rev. C 85 (2012) 044608.

\bibitem{Ismail} M. Ismail, A. Adel, Phys. Rev. C 89 (2014) 034617.

\bibitem{Seif} W. M. Seif, J. Phys. G: Nucl. Part. phys. 40 (2013) 105102.

\bibitem{Qian1} Yibin Qian, Zhongzhou Ren, Nucl. Phys. A 852 (2011) 82.

\bibitem{Qian2} Yibin Qian, Zhongzhou Ren, Dongdong Ni, Nucl. Phys. A 866 (2011) 1.

\bibitem{Qian3} Yibin Qian, Zhongzhou Ren, Phys. Rev. C 84 (2011) 064307.

\bibitem{Angeli} I. Angeli, K. P. Marinova, At. Data Nucl. Data Tables 99 (2013) 69.

\bibitem{Gurvitz} S. A. Gurvitz, P. B. Semmes, W. Nazarewicz, T. Vertse, Phys. Rev. A 69 (2004) 042705.

\bibitem{Wildermuth} K. Wildermuth, Y. C. Tang, {\sl A Unified Theory of the Nucleus} Academic Press, New York, 1997.

\bibitem{Bertsch} G. F. Bertsch, J. Borysowicz, H. Mcmanus, W. G. Love, Nucl. Phys. A 284 (1977) 399.

\bibitem{Satchler} G. R. Satchler, W. G. Love, Phys. Rep. 55 (1979) 183.

\bibitem{Ni2} Dongdong Ni, Zhongzhou Ren, Tiekuang Dong, Yibin Qian, Phys. Rev. C 87 (2013) 024310.

\bibitem{Moller} P. M\"{o}ller, J. R. Nix, W. D. Myers, W. J. Swiatecki, At. Data Nucl. Data Tables 59 (1995) 185.

\bibitem{Walecka} J. D. Walecka, {\sl Theoretical Nuclear Physics and Subnuclear Physics} Oxford University, Oxford, 1995, p. 11.

\bibitem{Hodgson} P. E. Hodgson, E. B\v{e}t\'{a}k, Phys. Rep. 374 (2003) 1.

\bibitem{Qian4} Yibin Qian, Zhongzhou Ren, Sci. China-Phys. Mech. Astron. 56 (2013) 1520.

\bibitem{AME} G. Audi, F. G. Kondev, M. Wang, B. Pfeiffer, X. Sun, J. Blachot, M. MacCormick, Chin. Phys. C 36 (2012) 1157.

\bibitem{NNDC} National Nuclear Data Center, Brookhaven National Laboratory
[http://www.nndc.bnl.gov].

\end{thebibliography}

%% else use the following coding to input the bibitems directly in the
%% TeX file.

\begin{table}[H]
\centering \caption{\label{tab1} Comparison of calculated $\alpha$
decay half-lives based on the corresponding measured charge radii
with available experimental values and other theoretical results for
long-lived $\alpha$-decaying nuclei ($T_{1/2}$ in years), including
the improved or new data about $^{146}$Sm, $^{151}$Eu and so on. The
last two columns denote the calculations within the UMADAC model
\cite{Denisov2} and the analytic formulas given by Royer
\cite{Royer2}. Predicted half-lives for hitherto undetected $\alpha$
emitters in nature are provided as well.} \vspace{0.5cm} \small
\begin{tabular}{lp{1.1cm}cp{0.6cm}cp{1.6cm}p{1.6cm}p{1.6cm}}
\hline Decay& $Q$(MeV) & $R^{\mathrm{expt}}$(fm) & $r_0$ &
$T_{1/2}^{\mathrm{expt}}$ &
 $T_{1/2}^{\mathrm{calc}}$ &
 $T_{1/2}^{\mathrm{\small UMADAC}}$ &
 $T_{1/2}^{\mathrm{form}}$\\
\hline
$^{144}$Nd$\rightarrow$$^{140}$Ce& 1.905 & 4.88 & 1.113 & 2.29$\pm$0.16$\times10^{15}$& 3.14$\times10^{15}$& 1.36$\times10^{16}$& 3.32$\times10^{15}$\\
$^{146}$Sm$\rightarrow$$^{142}$Nd& 2.529 & 4.91 & 1.118 & 6.8$\pm$0.7$\times10^{7}$& 6.9$\times10^{7}$& 2.0$\times10^{8}$& 1.0$\times10^{8}$\\
$^{147}$Sm$\rightarrow$$^{143}$Nd& 2.311 & 4.93 & 1.123 & 1.06$\pm$0.02$\times10^{11}$& 1.32$\times10^{11}$& 1.00$\times10^{12}$& 2.51$\times10^{11}$\\
$^{148}$Sm$\rightarrow$$^{144}$Nd& 1.986 & 4.94 & 1.120 & 7.00$\pm$2.00$\times10^{15}$& 7.98$\times10^{15}$& 4.48$\times10^{16}$& 8.36$\times10^{15}$\\
$^{151}$Eu$\rightarrow$$^{147}$Pm& 1.9489 & 4.99 & 1.075 & 4.62$\pm$1.63$\times10^{18}$& 3.71$\times10^{18}$& 5.95$\times10^{18}$& 1.28$\times10^{19}$\\
$^{152}$Gd$\rightarrow$$^{148}$Sm& 2.205 & 5.00 & 1.075 & 1.08$\pm$0.08$\times10^{14}$& 1.99$\times10^{14}$& 5.02$\times10^{14}$& 1.18$\times10^{14}$\\
$^{180}$W$\rightarrow$$^{176}$Hf& 2.516 & 5.33 & 1.082 & 1.1$^{+0.9}_{-0.5}$$\times10^{18}$& 7.18$\times10^{17}$& 2.79$\times10^{18}$& 2.95$\times10^{17}$\\
$^{186}$Os$\rightarrow$$^{182}$W& 2.822 & 5.36 & 1.085& 2.0$\pm$1.1$\times10^{15}$& 1.34$\times10^{15}$& 4.59$\times10^{15}$& 5.99$\times10^{14}$\\
$^{190}$Pt$\rightarrow$$^{186}$Os& 3.243 & 5.39 & 1.097 & 6.50$\pm$0.30$\times10^{11}$& 3.51$\times10^{11}$& 1.31$\times10^{12}$& 2.04$\times10^{11}$\\
$^{209}$Bi$\rightarrow$$^{205}$Tl& 3.137 & 5.48 & 1.129 & 2.03$\pm$0.08$\times10^{19}$& 1.78$\times10^{19}$& 2.96$\times10^{20}$& 3.21$\times10^{19}$\\
$^{244}$Pu$\rightarrow$$^{240}$U& 4.666 & 5.87 & 1.072 & 1.007$\pm$0.004$\times10^{8}$& 1.498$\times10^{8}$& 4.459$\times10^{8}$& 1.295$\times10^{8}$\\
$^{142}$Ce$\rightarrow$$^{138}$Ba& 1.310 & 4.84 & 1.108 & $>5\times10^{16}$& 4.02$\times10^{27}$& 3.43$\times10^{28}$& 2.37$\times10^{27}$\\
$^{145}$Nd$\rightarrow$$^{141}$Ce& 1.578 & 4.93 & 1.129 & & 6.00$\times10^{22}$& 1.07$\times10^{24}$& 1.60$\times10^{23}$\\
$^{149}$Sm$\rightarrow$$^{145}$Nd& 1.870 & 4.95 & 1.076 & $>2\times10^{15}$& 5.84$\times10^{18}$& 3.06$\times10^{19}$& 6.52$\times10^{18}$\\
$^{156}$Dy$\rightarrow$$^{152}$Gd& 1.758 & 5.08 & 1.072 & $>1.0\times10^{15}$& 5.98$\times10^{24}$& 2.81$\times10^{25}$& 1.88$\times10^{24}$\\
$^{162}$Er$\rightarrow$$^{158}$Dy& 1.645 & 5.18 & 1.061 & $>1.4\times10^{14}$& 4.56$\times10^{29}$& 1.80$\times10^{30}$& 9.23$\times10^{28}$\\
$^{164}$Er$\rightarrow$$^{160}$Dy& 1.304 & 5.20 & 1.059 & & 2.22$\times10^{40}$& 1.69$\times10^{41}$& 2.05$\times10^{39}$\\
$^{168}$Yb$\rightarrow$$^{164}$Er& 1.950 & 5.24 & 1.069 & $>1.3\times10^{14}$& 2.92$\times10^{24}$& 1.13$\times10^{25}$& 8.05$\times10^{23}$\\
$^{174}$Hf$\rightarrow$$^{170}$Yb& 2.559 & 5.29 & 1.070 & 2.00$\pm$0.40$\times10^{15}$& 4.43$\times10^{15}$& 1.07$\times10^{16}$& 2.02$\times10^{15}$\\
$^{176}$Hf$\rightarrow$$^{172}$Yb& 2.258 & 5.30 & 1.073 & & 2.25$\times10^{20}$& 7.76$\times10^{20}$& 7.67$\times10^{19}$\\
$^{178}$Hf$\rightarrow$$^{174}$Yb& 2.083 & 5.31 & 1.078 & & 3.34$\times10^{23}$& 1.59$\times10^{24}$& 9.54$\times10^{22}$\\
$^{182}$W$\rightarrow$$^{178}$Hf& 1.774 & 5.34 & 1.083 & & 2.66$\times10^{32}$& 2.82$\times10^{33}$& 3.97$\times10^{31}$\\
$^{184}$Os$\rightarrow$$^{180}$W& 2.957 & 5.35 & 1.085 & & 2.85$\times10^{13}$& 9.46$\times10^{13}$& 1.44$\times10^{13}$\\
$^{188}$Os$\rightarrow$$^{184}$W& 2.143 & 5.37 & 1.092 & & 1.02$\times10^{26}$& 8.17$\times10^{26}$& 2.27$\times10^{25}$\\
$^{192}$Pt$\rightarrow$$^{188}$Os& 2.418 & 5.40 & 1.103 & $>6\times10^{16}$& 5.09$\times10^{22}$& 4.72$\times10^{23}$& 1.46$\times10^{22}$\\
$^{196}$Hg$\rightarrow$$^{192}$Pt& 2.041 & 5.42 & 1.152 & $>2.5\times10^{18}$& 3.51$\times10^{31}$& 1.16$\times10^{33}$& 1.14$\times10^{31}$\\
$^{204}$Pb$\rightarrow$$^{200}$Hg& 1.9695 & 5.46 & 1.143 & $>1.4\times10^{20}$& 1.70$\times10^{35}$& 7.94$\times10^{36}$& 2.82$\times10^{34}$\\
\hline
\end{tabular}
\end{table}

\begin{table}
\centering \caption{\label{tab2} Calculated $\alpha$ decay
half-lives in the decay chain from the new nuclide $^{294}$117,
compared with the experimental values and other theoretical model
calculations (the UMADAC model \cite{Denisov2} and the analytical
formulas \cite{Royer2}). The measured data, i.e., the decay energies
$Q$ and the half-lives $T_{1/2}^{\mathrm{expt}}$, are obtained from
the very recent experiment \cite{Khuya}.} \vspace{0.5cm}
\begin{tabular}{lccccc}
\hline Nucleus& $Q$(MeV) & $T_{1/2}^{\mathrm{expt}}$ &
$T_{1/2}^{\mathrm{calc}}$ & $T_{1/2}^{\mathrm{UMADAC}}$&
$T_{1/2}^{\mathrm{form}}$\\
\hline
$^{294}$117 & 11.20(4) & 51$^{+94}_{-20}$ ms & 22-35 ms & 119-190 ms & 33-53 ms  \\
$^{290}$115 & 10.45(4) & 1.3$^{+2.3}_{-0.5}$ s & 0.4-0.7 s & 3.3-5.6 s & 0.7-1.1 s\\
$^{286}$113 & 9.4(3) & 2.9$^{+5.3}_{-1.1}$ s & 12.0-943.3 s & 102.5-9468.6 s & 19.9-1567.7 s\\
$^{282}$Rg & 9.18(3) & 3.1$^{+5.7}_{-1.2}$ min & 1.5-2.3 min & 11.4-17.9 min & 2.7-4.2 min \\
$^{278}$Mt & 9.59(3) & 3.6$^{+6.5}_{-1.4}$ s & 0.8-1.3 s & 4.3-6.5 s & 1.9-2.8 s \\
$^{274}$Bh & 8.97(3) & 30$^{+54}_{-12}$ s & 12-18 s & 59-93 s & 27-43 s \\
$^{270}$Db & 8.02(3) & 1.0$^{+1.9}_{-0.4}$ h & 1.0-1.6 h & 5.7-9.6 h & 2.4-3.9 h \\
\hline
\end{tabular}
\end{table}

\end{document}